\documentclass[e-only,10pt,reqno]{ofj}

\usepackage{tensorNotation}
\usepackage{finiteVolumeNotation}

\usepackage{setspace}
\usepackage{lineno}
\usepackage{color}
\usepackage{units}
\usepackage[dvipsnames,svgnames,x11names]{xcolor}
\usepackage{orcidlink}
\usepackage{cite}
\usepackage{subcaption}
\usepackage{graphicx}
\usepackage{units}

\let\paragraph\undefined

\copyrightinfo{2022}{The Authors. This work is an open access article under the {\color{blue}{\href{https://creativecommons.org/licenses/by-sa/4.0/}{CC BY-SA 4.0}}} license}

\OpenFOAMversions{FOAM-extend 4.1}
\Repository{https://bitbucket.org/C\_Habes/lcs4foam}

\DOI{Preprint submitted to OpenFOAM Journal} 

\begin{document}


\title[lcs4Foam]{lcs4Foam -- An OpenFOAM Function Object to\\ Compute Lagrangian Coherent Structures}




\author{C.\ Habes$^{1,*}$}

\author{A.\ Kameke$^{2,*}$}

\author{M. E.\ Fadeli$^1$}

\author{H.\ Marschall$^1$}
\address{$^1$Mathematical Modeling and Analysis, Technical University of Darmstadt, 64287 Darmstadt, Germany}
\email{constantin.habes@tu-darmstadt.de, elwardi.fadeli@tu-darmstadt.de, marschall@mma.tu-darmstadt.de}

\address{$^2$Department of Mechanical Engineering and Production Management, Hamburg University of Applied Sciences, 20099 Hamburg, Germany}
\email{alexandra.vonkameke@haw-hamburg.de}


\begin{abstract}
    To facilitate the understanding and to quantitatively assess the material transport in fluids, a modern characterisation method has emerged in fluid dynamics in the last decades footed in dynamical systems theory. It allows to examine the most influential material lines which are called Lagrangian Coherent Structures (LCS) and order the material transport into dynamically distinct regions at large scales which resist diffusion or mixing. LCS reveal the robust skeleton of material surfaces and are essential to assess material transport in time-dependent flows quantitatively. Candidates of LCS can be estimated and visualised from finite-time stretching and folding fields by calculating the Finite-Time Lyapunov Exponents (FTLE).
\medskip

\noindent
In this contribution, we provide an OpenFOAM function object to compute FTLE during CFD simulation. This enables the OpenFOAM community to assess the geometry of the material transport in any flow quantitatively on-the-fly using principally any OpenFOAM flow solver.
\end{abstract}

\date{\today}

\dedicatory{}

\maketitle



\section{Introduction}\label{sec:intro}

    Material transport and mixing in fluids is enhanced by advection. This advection is usually described mathematically in an Eulerian view by a time-dependent velocity field $\mathbf{u}(\mathbf{u},\,t)$. With this Eulerian description, numerous important fluid mechanical characteristics can be derived and assessed. For instance, a higher Reynolds number (higher velocities) typically will go along with better overall mixing. However, such intuition might be misleading as has been shown for example in \cite{kameke_how_2019} studying a rising bubble. Here, a coherent structure has been found to arise for intermediate Reynolds numbers, which causes material to move together and locally hinders mixing and increases residence times in the vicinity of the bubble rear. The example shows: a closer look at the coherent structures is necessary to evaluate the details of the material transport in the specific flow situation. Lagrangian Coherent Structures (LCS) are often observable in fluid flows due to the shape that passive tracers take on, e.g.\ plankton in the ocean \cite{huhn_impact_2012} or dissolved oxygen in the wake behind a rising bubble.

That the classical Eulerian view on advection is not optimal for addressing these issues was first noted in oceanography and atmospheric science \cite{dovidio_mixing_2004, weiss_transport_2008}. The transport analysis was therefore started from its roots, the Lagrangian view, where the observer travels on the fluid parcels rather than watching them move by (Eulerian frame). The Lagrangian analysis thus considers the trajectories of individual fluid parcels and allows to draw conclusions on the transport from their evaluation. Nowadays, computational and theoretical advances allow for the calculation and analysis of the time-dependent dynamical system that governs material transport.
The underlying ideas for Lagrangian analysis stem from dynamical systems theory. In time-independent incompressible velocity fields, the dynamical system is the velocity field itself and the streamlines of the velocity field coincide with the trajectories of the fluid parcels. As such, trivially, structures in the velocity fields represent governing structures for the material that is transported (as long as molecular diffusion is comparably low and negligible) \cite{neufeld_chemical_2009}. In this setting, unstable and stable manifolds divide the flow into different subdomains that move coherently (together) \cite{balasuriya_generalized_2018}.
For time-dependent flows however, the instantaneous streamlines and the trajectories of the fluid parcels do not coincide. It is thus a misleading habit to draw any conclusion about the material transport from the streamlines or any other material lines of the mean velocity field of a fluid flow. The resulting transport structures might have no relevance for the real dynamical system at all.

To obtain the lines that govern material transport in time-dependent flows the Lagrangian Coherent Structures are calculated from the trajectories of particles evaluated in the time-dependent velocity field $\mathbf{u} = \mathbf{u}(\mathbf{x},\,t)$. LCS are those material lines and surfaces that separate regions of particles with very different fates or history for the time interval under consideration. Several different approaches to evaluate LCS have been developed during the last years \cite{hadjighasem_critical_2017}.

With this contribution we introduce an OpenFOAM function object that calculates the three dimensional Finite Time Lyapunov Exponents (FTLE) on-the-fly based on the general purpose numerical library \emph{libcfd2lcs} \cite{justin_finn_libcfd2lcs_2017} with the main computational details explained in \cite{finn_integrated_2013}. The ridges in the FTLE-field are then candidates for LCS and can be assumed to coincide with LCS if some further conditions are met \cite{farazmand_computing_2012}. However, as also pointed out in \cite{justin_finn_libcfd2lcs_2017}, these additional conditions are hard to evaluate in 3D and thus the FTLE-field will be viewed as an approximate representation of the 3D LCS. The details about the calculation of the FTLE-field and the underlying mathematical foundation are set out in Section \ref{sec:theoback}.
    
\section{Theoretical background of LCS calculations}\label{sec:theoback}

    From time-resolved CFD simulations, the time-dependent velocity field $\mathbf{u}(\mathbf{x}, t)$ is known in space and time. From this information the fluid parcel or passive particle trajectories 
\begin{align}\label{eq:parcel_trajectories}
    \mathbf{x}(\mathbf{x_0}, t) = \mathbf{x}_{{0}} + \int_{t_{0}}^{t} \mathbf{u}(\mathbf{x}(\tau), \tau) d \tau
\end{align}
can be calculated, where $\mathbf{x}_0$ is the starting point of a trajectory in 3D space at a starting time $t_0$. Note, that each trajectory is now labelled by its start location in space and time. If a set of initially close passive particles is released at the same time the distances between them change over time due to the fluid motion. Passive particles initially forming a tiny sphere will undergo a linear deformation towards an ellipse for short times as would occur in a solid body under stress before it breaks. Certainly, in a fluid, the deformation will progress, and non-linear higher-order terms will play a role in causing stretching and folding which is crucial for mixing. However, as a first approximation and for short times these higher-order terms are neglected for the analysis of the deformation. If we consider infinitesimal spheres of initially close particles around all mesh cell centres of our simulation starting at the same initial time $t_0$, we obtain a set of different ellipsoids. All these ellipsoids have differently stretched and contracted principal axes which point in different directions at a slightly later time $t_1$. The principal axes of each ellipse denominate the final directions of maximal stretching (major axis) and maximal contraction (minor axis) of the initially spherical particle blob. The stretching factor $S$ is the length of the major axis of the final ellipse divided by the initial radius of the sphere. If this stretching factor at each initial grid point is plotted, a 3D stretching field results revealing the regions at which stretching and thus particle separation for the time interval of interest $[t_0,t_1]$ is largest due to the local flow conditions. Normally, the scaled logarithm of this stretching factor, defined by 
\begin{align}\label{eq:FTLE1}
    \sigma_{t_{0}}^{t_{1}}\left(\mathbf{x}_{0}, t_{0}\right)=\frac{1}{\left|t_{1}-t_{0}\right|} \log (S) \;,
\end{align}
is plotted. This scaled logarithmic stretching factor is called the Finite-Time Lyapunov Exponent \cite{shadden_definition_2005}. Connected areas or lines of large FTLE values characterise the fluid transport as these denote the areas or lines along which deformation and thus particle separation is largest. All these geometrical considerations have their mathematical counterparts. The stretching factor as described is the square root of the maximal eigenvalue of the right Cauchy-Green deformation tensor $\mathbf{C}_{t_{0}}^{t_{1}}$. This tensor can be calculated for every mesh cell as envisioned above for the ellipsoid. As its name reveals it includes all the information about the deformation of the fluid masses at this point for the short time interval $t_1 -t_0$, and notably it is an objective tensor such that high stretching values and candidates for LCS derived from it will persist regardless of the motion of the observer (invariant to a time-dependent translation and rotation of the coordinate system of the observer) \cite{haller_lagrangian_2015,shadden_definition_2005}.

The governing ordinary differential equation (ODE) for the evolution of a fluid parcel or a passive particle reads
\begin{align}\label{eqn:parcel_ODE}
    \dot{\mathbf{x}}=\mathbf{u}(\mathbf{x}(t), t) \;.
\end{align}
Therefore, the infinitesimal separation $\mathbf{\gamma } = \mathbf{x}-\mathbf{x}^*$ of the passive particle, imagined in the centre of a infinitesimal sphere, to a particle on the surface of the sphere will be governed by the ODE 
\begin{align}\label{eqn:separationODE}
\mathbf{\delta \dot{\mathbf{x}}} = \nabla  \mathbf{u} \mathbf{\gamma} \;.
\end{align}
The solution of this ODE is an exponential function, which explains why the FTLE is defined as the logarithm of the stretching factor.

To analyse the stretching during short but finite time intervals, particles distributed on a mesh are advected with the flow from an initial time $t_{0}$ over the time interval $T=\left|t_{1}-t_{0}\right|$ to $t_{1}$. From the integral version of the governing ODE (Eq.\ \ref{eqn:parcel_ODE}) we obtain the definition of the flow map, $\Phi_{t_{0}}^{t_{1}}$, which maps all the particles from their initial positions onto their final positions at time $t_{1}$, viz.\
\begin{equation}\label{eqn:Flow_map}
\Phi_{t_{0}}^{t_{1}}: \mathbb{R}^{n} \rightarrow \mathbb{R}^{n} ; \quad \mathbf{x}_{0} \mapsto \mathbf{x}_{0} + \int_{t_{0}}^{t_{1}} \mathbf{u}(\mathbf{x}(\tau), \tau) d \tau \;.
\end{equation}

To obtain the separation of two initially close particles after this time interval a Taylor series 
\begin{align}
    \delta \mathbf{x}(t_{1}) = \Phi_{t_{0}}^{t_{1}}(\mathbf{x}_{0})-\Phi_{t_{0}}^{t_{1}}(\mathbf{x_0 + \delta x(t_0)}) = \mathbf{D} \boldsymbol{\Phi}_{t_{0}}^{t_{1}}(\mathbf{x}_{0}, t_{0}) \delta x(t_0) + \mathcal{O}(| \delta x(t_0)^2|)
\end{align}
around the initial position can be employed. Where $\mathbf{D} \boldsymbol{\Phi}_{t_{0}}^{t_{1}}(\mathbf{x}_{0}, t_{0})$ is the gradient (Jacobian) of the flow map with regard to the initial separation and is also the normalised fundamental matrix solution of the equation of variations above (Eq.\ \ref{eqn:separationODE}) \cite{haller_lagrangian_2015}. Therefore, the particle separation at time $t_1$ can be written as
\begin{align}\label{eqn:Particle_separation_after_flow_map}
\left\|\delta \mathbf{x}\left(t_{1}\right)\right\|=\sqrt{\left\langle\delta \mathbf{x}(t_0),\left[\mathbf{D} \boldsymbol{\Phi}_{t_{0}}^{t_{1}}\left(\mathbf{x}_{0}, t_{0}\right)\right]^{*}\left[\mathbf{D} \boldsymbol{\Phi}_{t_{0}}^{t_{1}}\left(\mathbf{x}_{0}, t_{0}\right)\right] \delta \mathbf{x}(t_0)\right\rangle}.
\end{align}
The right Cauchy-Green deformation tensor is then defined as
\begin{align}\label{eqn:Right_Cauchy-Green_deformation_tensor} 
\mathbf{C}_{t_{0}}^{t_{1}}\left(\mathbf{x}_{0}, t_{0}\right)=\left[\mathbf{D} \Phi_{t_{0}}^{t_{1}}\left(\mathbf{x}_{0}, t_{0}\right)\right]^{*}\left[\mathbf{D} \Phi_{t_{0}}^{t_{1}}\left(\mathbf{x}_{0}, t_{0}\right)\right] \;.
\end{align} 
In this way the Finite-Time Lyapunov Exponent $\sigma_{t_{0}}^{t_{1}}$ for the time interval $t_{0}$ to $t_{1}$ can now be defined on the basis of this tensor in a more thorough, mathematical way. Therefore, it is now defined by
\begin{align}\label{eqn:FTLE}
\sigma_{t_{0}}^{t_{1}}\left(\mathbf{x}_{0}, t_{0}\right)=\frac{1}{\left|t_{1}-t_{0}\right|} \log \sqrt{\lambda_{\max }\left(\mathbf{C}_{t_{0}}^{t_{1}}\left(\mathbf{x}_{0}, t_{0}\right)\right)} \;.
\end{align}
Here $\lambda_{\max}$ is the maximum eigenvalue of the right Cauchy-Green deformation tensor and can be calculated using standard solvers. In the picture of the small ellipsoid, the square root of the eigenvalue is just the above stretching rate $S$.
    
\section{Computational details}\label{sec:Computational_details}
    The computation of flow maps within \textit{libcfd2lcs} is described thoroughly in \cite{finn_integrated_2013}. The following section presents a brief overview of how the computation is done in practice and which different timescales play a role in the calculations. Hereafter, we describe the structure and functionality of the newly developed function object. We will focus on how the function object acts as an interface between OpenFOAM and \textit{libcfd2lcs}, how parallelisation is ensured and what has to be considered for the output of the generated data.
    \subsection{Numerical flow map computation in libcfd2lcs}\label{sec:Numerical_flow_map_computation}
        \textit{libcfd2lcs} is able to calculate both forward-time and backward-time FTLE fields. However, it uses two very different approaches for calculating the respective flow maps. The general approach used for the computation of the forward time flow-map $\Phi_{t_{0}}^{t_{0}+T}$ and the resulting forward-time FTLE field is very straightforward. A set of tracer particles is initialised on a grid with spacing $\Delta x_{lcs}$ by setting each initial tracer coordinate to the cell centre coordinate of a corresponding mesh cell. Then the flow map at each cell centre is computed by passively advecting these tracers with the flow, which mathematically corresponds to an integration of equation 
\begin{align}\label{eqn:Lagrangian_advection}
\frac{d \mathbf{x}}{d t}=\mathbf{u}(\mathbf{x}, t)
\end{align}
over the time interval $T$. Numerically this integration is done by utilising Runge-Kutta methods, with step size $\Delta t_{lcs}$.
The time and space dependent velocity field $\mathbf{u}(\mathbf{x}, t)$ results from the specific fluid simulation under consideration and is passed to \textit{libcfd2lcs} after each simulation time step $\Delta t_{sim}$ (see Section \ref{sec:Structure_of_the_function_object}). In order to save the flow map $\Phi_{t_{0}}^{t_{0}+T}$, the location of each particle after the integration is stored at its initial position.

As the evaluation of FTLE fields, indicating LCS candidates, is mainly relevant for time-dependent flows, it is often important to animate their evolution. At first glance, this would mean that a sequence of large particle sets would have to be integrated, requiring a great amount of computation. This problem is solved using a method developed by Brunton and Rowley \cite{brunton_fast_2010}. With this method a flow map of the interval $T$ can be constructed from a sequence of $k$ flow maps over a smaller interval $h$, where $T=kh$. Following the notation of \cite{brunton_fast_2010} this can be expressed as 
\begin{equation}\label{eqn:Fwd_time_flow_map_construction}
    \Phi_{t_{0}}^{t_{0}+T}=\Phi_{t_{0}+(k-1)h}^{t_{0}+kh} \circ \cdots \circ \Phi_{t_{0}+h}^{t_{0}+2h} \circ \Phi_{t_{0}}^{t_{0}+h} 
    \;.
\end{equation}
In practical terms, this means that the particle grid is reinitialised for every new time interval $h$ after which they are advected again with the flow. Then the sub-step flow map is stored and the complete flow map is constructed when all needed sub-step flow maps are available. It is important to note that since a discrete particle grid is used for the sub-step flow map computation, interpolation of the sub-step flow maps is needed in order to match the trajectories at different timelevels when reconstructing the flow map $\Phi_{t_{0}}^{t_{0}+T}$ (see \cite{brunton_fast_2010} for more details).

A different approach for constructing the backward-time flow maps is used. This is due to the fact that using the Lagrangian approach would require to store all computed velocity fields in the subset interval $h$ before the integration of the tracers from $t_{0}+h$ to $t_{0}$ could be done backward in time. Although this already includes Brunton's and Rowley's method for the flow map construction, the Lagrangian approach would be "cumbersome and resource intensive" \cite[p. 4]{finn_integrated_2013}. Therefore, \textit{libcfd2lcs} uses an Eulerian approach for the flow map computation proposed by Leung \cite{leung_eulerian_2011}. In contrast to the forward-time flow map, the backward-time flow map $\Phi_{t_{0}+T}^{t_{0}}$ describes for each grid point where a particle, that is at that point at time $t_{0}+T$, originally was at time $t_{0}$. With Leung's Eulerian approach this backward-time flow map at time $t_{0}+T$ is computed by initialising a vector field $\mathbf{\Psi}(\mathbf{x}, t_{0})$ on a grid with the cell centre coordinates at time  $t_{0}$. The advection of this so called "takeoff coordinate field" in an Eulerian reference frame is then described by the level set equation
\begin{equation}\label{eqn:Level_set}
    \frac{\partial \mathbf{\Psi}(\mathbf{x}, t)}{\partial t}+(\mathbf{u} \cdot \nabla) \mathbf{\Psi}(\mathbf{x}, t)=\mathbf{0}
    \;.
\end{equation}
Solving this equation over the time Interval $[t_{0}, t_{0}+T]$ in forward time gives $\mathbf{\Psi}(\mathbf{x}, t_{0}+T)$, which represents the takeoff coordinates of a Lagrangian particle at $t_{0}$ reaching $\mathbf{x}$ at time $t_{0}+T$. Thus, the backward-time flow map $\Phi_{t_{0}+T}^{t_{0}}$ is equivalent to $\mathbf{\Psi}(\mathbf{x}, t_{0}+T)$. \textit{libcfd2lcs} solves equation (\ref{eqn:Level_set}) by  using a semi-Lagrangian advection approach with the time step size 
\begin{equation}\label{eqn:CFL_time_step}
    \Delta t_{lcs} = \frac{c_{cfl} \, \Delta x_{lcs}}{\mathbf{u}(\mathbf{x}, t)}
\end{equation}
of this procedure being restricted by the CFL condition $c_{cfl} < 1$ (see \cite{finn_integrated_2013} and \cite{leung_eulerian_2011} for more details).
Furthermore, Brunton's and Rowley's flow map construction method is also applied to the backward-time flow maps computed with the Eulerian method. Hence, the takeoff coordinate field is reinitialised after every sub-step time interval $h$ and the backward-time flow map
\begin{equation}\label{eqn:Bkwd_time_flow_map_construction}
    \Phi_{t_{0}+T}^{t_{0}}=\Phi_{t_{0}+h}^{t_{0}}\circ \Phi_{t_{0}+2h}^{t_{0}+h} \circ \cdots \circ \Phi_{t_{0}+kh}^{t_{0}+(k-1)h} 
\end{equation}
is constructed form $k$ sub-step backward-time flow maps.

\begin{figure}
    \centering
    \includegraphics[width=\textwidth]{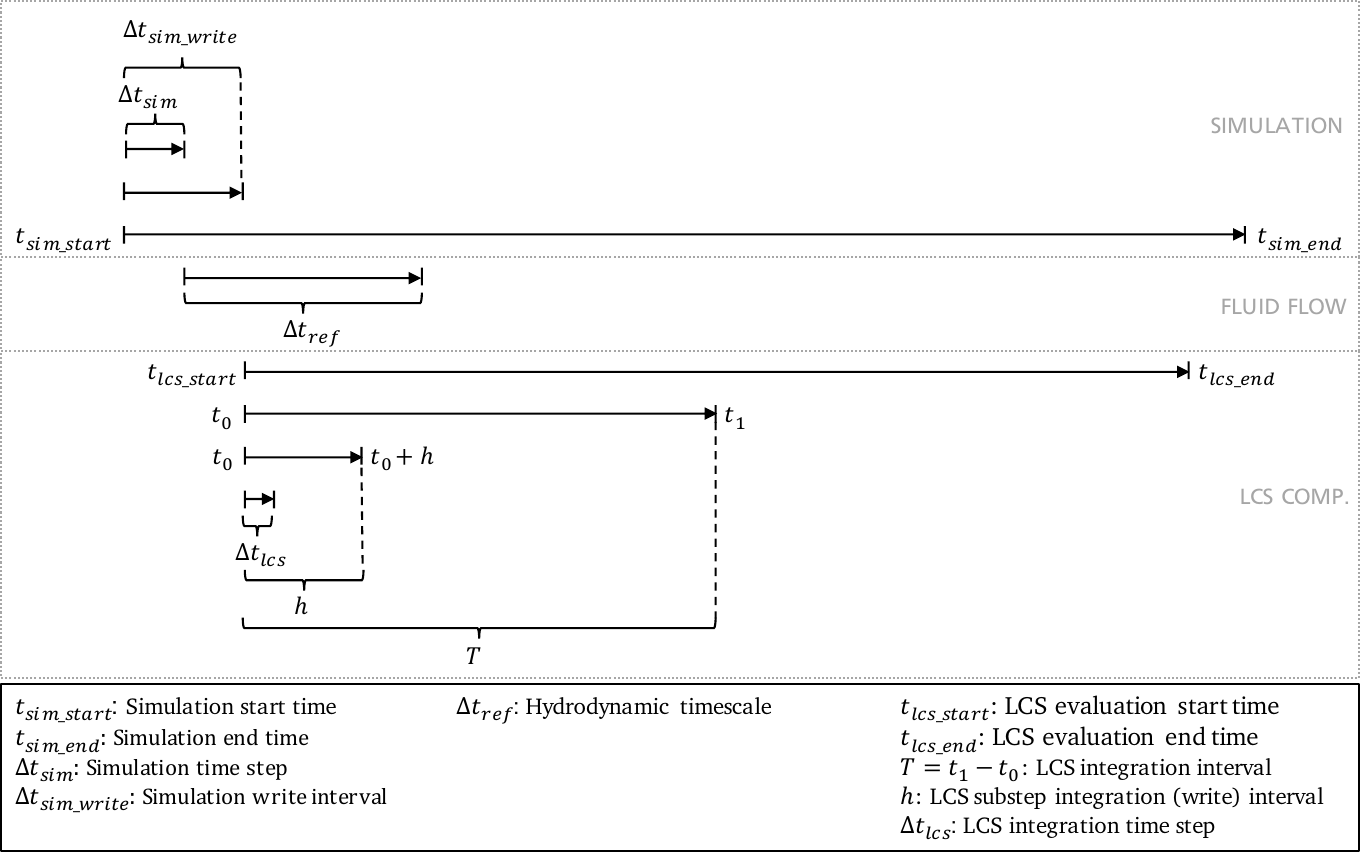}
    \caption{Different timescales relevant for the LCS evaluation}
    \label{fig:lcs_time_intervals}
\end{figure}

Since a lot of different timescales are relevant in the practical FTLE field computation described above, we try to differentiate and order them in the following, before describing the structure and functionality of the newly developed function object in the next section. The basis of the on-the-fly LCS evaluation is a parallel running simulation that provides the velocity fields. Here, three intervals are of interest (see Fig.\ \ref{fig:lcs_time_intervals}): the overall simulation time that spans from the simulation start time $t_{sim\_start}$ to the simulation end time $t_{sim\_end}$, the time step size of the simulation $\Delta t_{sim}$ and the write time interval of the simulation results $\Delta t_{sim\_write}$. The computed velocity fields represent a fluid flow for which a reference timescale $\Delta t_{ref}$ can be identified. This reference timescale characterises the dominant hydrodynamic timescale of the flow and is typically larger than the simulation time step size. In order to save computing resources the LCS evaluation of the simulated flow does not necessarily have to start and end at the same time as the simulation. Therefore, a separate start and end time for the LCS evaluation denoted as $t_{lcs\_start}$ and $t_{lcs\_end}$ can be defined (see Fig.\ \ref{fig:lcs_time_intervals}). During the LCS evaluation, a series of FTLE fields are computed. These FTLE fields are calculated from time $T$ flow maps, which themselves are calculated as described earlier in this section. This means storing and constructing the time $T$ flow maps from multiple sub-step flow maps after each LCS sub-step integration interval $h$. Calculating the sub-step flow maps in turn requires to numerically solve the equations (\ref{eqn:Lagrangian_advection}) or (\ref{eqn:Level_set}) using the finite time step $\Delta t_{lcs}$. While $\Delta t_{lcs}$ is set automatically according to equation (\ref{eqn:CFL_time_step}) and a specified CFL number, $T$ and $h$ have to be defined by the user. In order to detect all LCS candidates, $T$ is usually chosen to be larger than $\Delta t_{ref}$ of the investigated flow \cite{finn_integrated_2013}. With the aim of animating the evolution of the FTLE field, $h$ is typically set significantly smaller than $\Delta t_{ref}$ while being in the order of magnitude of $\Delta t_{sim\_write}$.

    \subsection{Structure and functionality of the function object}\label{sec:Structure_of_the_function_object}
        In general, function objects can be used to generate additional data at runtime of the simulation. In doing so, function objects can access data generated by the flow solver at runtime, which offers a great advantage over classical post-processing since it can only utilise the stored fields or logged information. The newly developed function object incorporates the functionalities of \textit{libcfd2lcs} into OpenFOAM at runtime while acting as an interface between both. This is achieved by processing the data generated by OpenFOAM and the subsequent exchange of this data via the \textit{libcfd2lcs} API (see \cite{justin_finn_libcfd2lcs_2017} for a detailed description of the \textit{libcfd2lcs} API).

The calculation of the flow maps, the calculation of the resulting FTLE fields and the subsequent saving of these fields is completely handled by \textit{libcfd2lcs}. The basic task of the function object is to pass the cell centre position vectors of the computational grid as well as the velocity field calculated by OpenFOAM to \textit{libcfd2lcs}. Due to the very strict data structure requirements of \textit{libcfd2lcs} this is not a trivial task. \textit{libcfd2lcs} can only use static rectlinear grids for the calculation of forward-time and backward-time flow maps and therefore needs the velocity fields on these grids. This means that the mesh and velocity data has to be globally organised in an $(i,\, j,\, k)$ structured format \cite[p. 6]{justin_finn_libcfd2lcs_2017}. Since the LCS evaluation should also be available for simulations on moving grids with general topology and adaptive grid refinement, the function object offers several possibilities to deal with this problem. 

In the simplest case, where the simulation mesh is already a static rectlinear mesh, the function object does not need to process the grid and velocity data, but can directly transfer it to \textit{libcfd2lcs} as basic C++ arrays. This is the preferred method when the flow domain can be represented by a static rectlinear mesh and e.g. immersed boundary methods are used. If a moving mesh, a mesh of general topology or adaptive mesh refinement is used for the simulation a different approach is needed in order to prepare the data for its use in \textit{libcfd2lcs}. Here, an additional static rectlinear mesh needs to be constructed in the preprocessing step, which can be done e.g. by using the \texttt{blockMesh} utility. This mesh has to contain the region for which the LCS diagnostic should be performed, meaning that it can cover the whole simulation domain as well as only a part of it. However, since \textit{libcfd2lcs} also requires boundary conditions for the FTLE field calculations, the boundary patches of the additional LCS mesh must be set accordingly. The user can choose between \texttt{empty}, \texttt{symmetryPlane}, \texttt{wedge}, \texttt{cyclic} and the generic \texttt{patch} patch types which the function objects translates into the corresponding \textit{libcfd2lcs} boundary types. Then, during runtime, the velocity fields are mapped from the simulation mesh of general topology to the static rectlinear LCS mesh, from which the data can again be transferred to \textit{libcfd2lcs} as basic C++ arrays. Although this implies that interpolation errors are made during the mapping process, the LCS evaluation is hardly affected by this. Haller showed in \cite{haller_lagrangian_2002} that LCS are very robust against errors in the velocity field. Also, the additional computational overhead due to the mapping can be neglected compared to the overhead caused by the flow map computations. The function object also implements a third approach in which no additional LCS mesh is needed. This approach utilises the ability to construct complex, moving mesh geometries out of simple unconnected mesh regions in OpenFOAM with the \texttt{oversetMesh} approach. Using this approach the function object can utilise any specified static rectlinear mesh region of the \texttt{oversetMesh} for the LCS evaluation, meaning that the background mesh as well as any other static rectlinear mesh region can be used. In doing so, the function object extracts the mesh and velocity data from the specified mesh region of the \texttt{oversetMesh} and passes it to \textit{libcfd2lcs} analogously to the previous approaches. Here the \texttt{overset} type patches are generally passed on as inlet or outlet, as they are treated the same by \textit{libcfd2lcs}.

\begin{figure}
    \centering
    \includegraphics[width=0.45\textwidth]{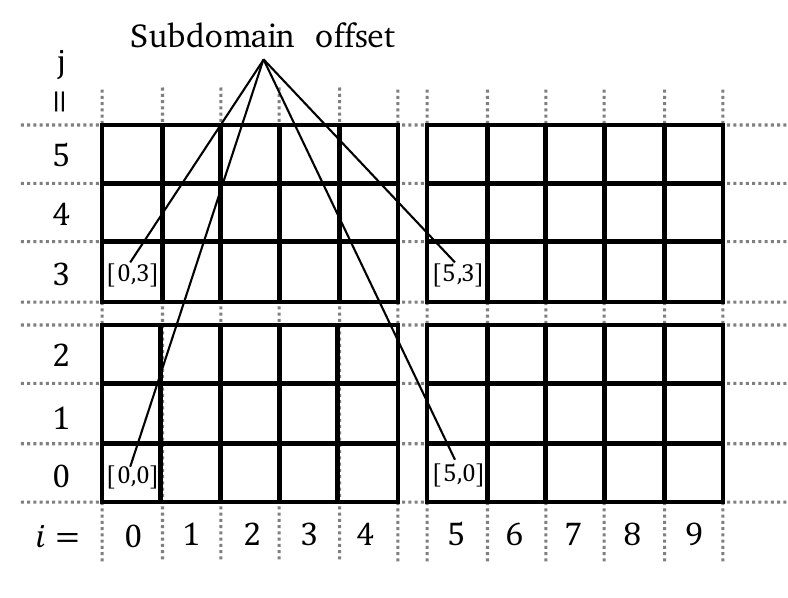}
    \caption{Subdomain offset of a $i,\, j,\, (k=0)$ structured rectlinear mesh}
    \label{fig:subdomain_offset}
 \end{figure}

As \textit{libcfd2lcs} also uses the domain decomposition approach and MPI for the parallelisation of the computations, the integration within the parallelisation of OpenFOAM is done in a straightforward manner. The local subdomains of the rectlinear LCS mesh and its velocity data are passed to \textit{libcfd2lcs} together with an offset, which describes the position of the cell data in the globally $(i,\, j,\, k)$ structured data array (see Fig.\ \ref{fig:subdomain_offset}). For the MPI communication, the same MPI communicator as used for OpenFOAM is shared with \textit{libcfd2lcs}. Therefore, the function object can be used for simulations running in parallel or serial. However, if the approach involving an additional LCS mesh is used, special attention is required for the domain decomposition in the preprocessing step. Here the simulation mesh, as well as the LCS mesh, must be cut along the same surfaces to make sure that the mapping of the velocity fields from one mesh to the other works properly.

As already mentioned, the data output of the flow-map and FTLE field data is completely handled by \textit{libcfd2lcs}. This is due to the fact that the data output interval defined by $h$ can differ from the solver write interval $\Delta t_{sim\_write}$ (see section \ref{sec:Numerical_flow_map_computation}). Therefore, the results generated by the function object are not stored in corresponding time directories but in a separate folder in the case directory called \texttt{cfd2lcs\_output}. Additionally, a directory named \texttt{cfd2lcs\_temp} is created inside of which all the sub-step data is stored. All data is stored in the Tecplot ASCII data file format (*.dat) and therefore can be visualised in ParaView when opened with its internal Tecplot reader or other common visualisation programs. In addition to this data, the computational overhead generated by the use of the function object with respect to the actual simulation is also output in the solver log file after each simulation time step. This enables the user to examine the computational costs of the LCS evaluation.

\section{Examples of usage}
    In this section a few examples are presented which are designed to show the functionality and capabilities of the function object. Therefore, example cases are presented in which only a rectlinear simulation mesh, a separate simulation and LCS mesh and a single \texttt{oversetMesh} are used.
    \subsection{Steady ABC flow}
        The Arnold-Beltrami-Childress (ABC) flow is an exact periodic solution of the Euler equations and is often used in the literature to verify LCS calculation methods. Therefore this case is also being reviewed here. The velocity field 
\begin{align} \label{eqn:ABC_vel_field}
\mathbf{u}=\nabla \times[-\Psi \mathbf{k}+\nabla \times(\Phi \mathbf{k})]
\end{align}
of the ABC flow can be described using 2 scalar potentials $\Psi$ and $\Phi$ \cite{sulman_leaving_2013} which themselves are defined as
\begin{align}\label{eqn:ABC_potential}
\begin{aligned}
&\Psi=-[C \sin (y)+B \cos (x)] \\
&\Phi=A[-x \cos (z)+y \sin (z)]-\Psi
\;.
\end{aligned}
\end{align}
In (\ref{eqn:ABC_vel_field}), $\mathbf{k}$ can be any unit vector but is commonly chosen to be the vertical unit vector. This leads to the three expressions of the velocity components 
\begin{align}\label{eqn:ABC_vel_comp}
\begin{aligned}
&u=A \sin (z)+C \cos (y) \\
&v=B \sin (x)+A \cos (z) \\
&w=C \sin (y)+B \cos (x) \;.
\end{aligned}
\end{align}

\begin{figure}[t]
\centering
\begin{subfigure}{0.25\textwidth}
    \includegraphics[width=\textwidth]{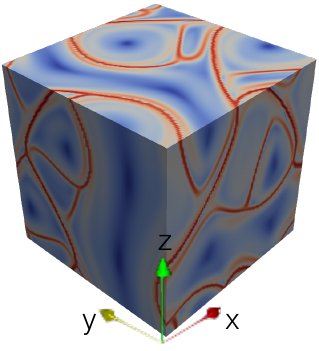}
    \caption{Forward-time FTLE field for the whole computational domain}
    \label{fig:1}
\end{subfigure}
\hspace{5mm}
\begin{subfigure}{0.25\textwidth}
    \includegraphics[width=\textwidth]{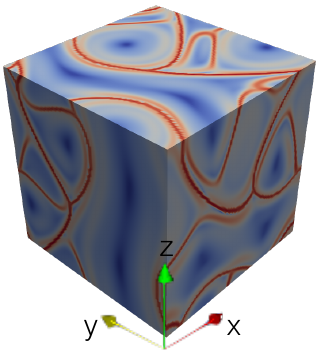}
    \caption{Backward-time FTLE field for the whole computational domain}
    \label{fig:2}
\end{subfigure}
\\ \vspace{4mm}
\begin{subfigure}{0.25\textwidth}
    \includegraphics[width=\textwidth]{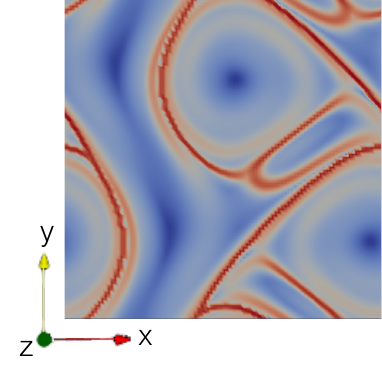}
    \caption{Forward-time field for a cross-section at $z=2\pi$}
    \label{fig:3}
\end{subfigure}
\hspace{5mm}
\begin{subfigure}{0.25\textwidth}
    \includegraphics[width=\textwidth]{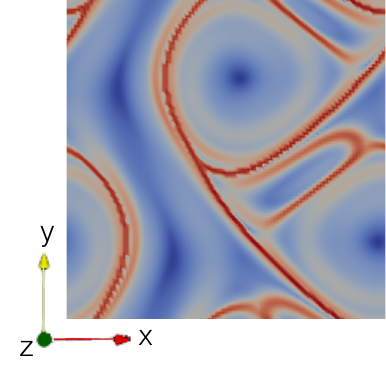}
    \caption{Backward-time field for a cross-section at $z=2\pi$}
    \label{fig:4}
\end{subfigure}
\\
\begin{subfigure}{0.35\textwidth}
    \includegraphics[width=\textwidth]{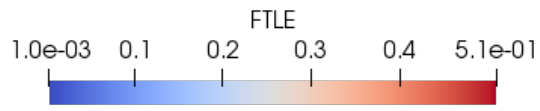}
\end{subfigure}
\caption{FTLE fields of a stationary ABC flow }
\label{fig:ABC_flow_results}
\end{figure}

The parameters $A$, $B$ and $C$ can be freely selected and influence the properties of the ABC flow. In order to create comparability with literature values, $A=0.5$, $B=0.8$, $C=0.8$ is chosen. In order to test the newly developed function object on this flow configuration a dedicated ABC flow OpenFOAM solver was written. This solver does not solve the Euler equations in the usual sense, but sets the velocity components on a given computational mesh according to (\ref{eqn:ABC_vel_comp}). Due to the periodicity of the flow solution, the dimensions of the computational mesh used in this case setup are specified as $x,y,z \in [0,2\pi]$ with a mesh size of $100\times100\times100$.
Since the described mesh is rectlinear no additional LCS mesh is used. Again for reasons of comparability, a LCS integration time of $T=10 \unit{s}$ is selected for the LCS evaluation. The results of the LCS evaluation, both in forward- and backward-time, can be seen in Figure \ref{fig:ABC_flow_results}. 

In these results the FTLE ridges,  which indicate the LCS candidates in the ABC flow, can be seen very clearly. Furthermore, the results agree very well with the results from \cite{sulman_leaving_2013}, both qualitatively and quantitatively, which suggests that the new function object calculates the FTLE ridges reliably.
    \subsection{Time dependent double gyre}
        Another frequently used flow for the verification of LCS computing algorithms is the time periodic Rayleigh-Bénard convection flow, or often called double gyre, proposed by Solomon and Gollub \cite{solomon_chaotic_1989}. The velocity field of this flow can be describe by using a stream function $\psi$
\begin{align}
\begin{aligned}
&u=-\frac{\partial \psi}{\partial y} \\
&v=\frac{\partial \psi}{\partial x}
\;.
\end{aligned}
\end{align}
Here $\psi$ is defined by 
\begin{align}
\psi(x, y, t)=A \sin (\pi f(x, t)) \sin (\pi y)
\end{align}
with
\begin{align}
\begin{aligned}
&f(x, t)=a(t) x^{2}+b(t) x \\
&a(t)=\epsilon \sin (\omega t) \\
&b(t)=1-2 \epsilon \sin (\omega t)
\end{aligned}
\end{align}
This leads to the expressions for two-dimensional velocity components
\begin{align}\label{eqn:double_gyre_vel_comp}
\begin{aligned}
&u=-\pi A \sin (\pi f(x)) \cos (\pi y) \\
&v=\pi A \cos (\pi f(x)) \sin (\pi y) \frac{\mathrm{d} f}{\mathrm{~d} x}
\;.
\end{aligned}
\end{align}
As the name double gyre suggests, this model defines the flow of two two-dimensional gyres enclosed in a rectangle which expand and contract periodically along the x-axis. Therefore, the periodic motion is controlled by $\epsilon$ if $\epsilon \neq 0$. Then $\epsilon$ describes approximately how far the line separating the gyres moves to the left or right from its centre position \cite{shadden_definition_2005}. Otherwise ($\epsilon=0$), no periodic motion is happening. Furthermore, $A$ specifies the magnitude of the velocity vectors and $\omega/2\pi$ determines the oscillation frequency of the gyres. 


\begin{figure}[t]
\centering
\begin{subfigure}{0.47\textwidth}
    \includegraphics[width=\textwidth]{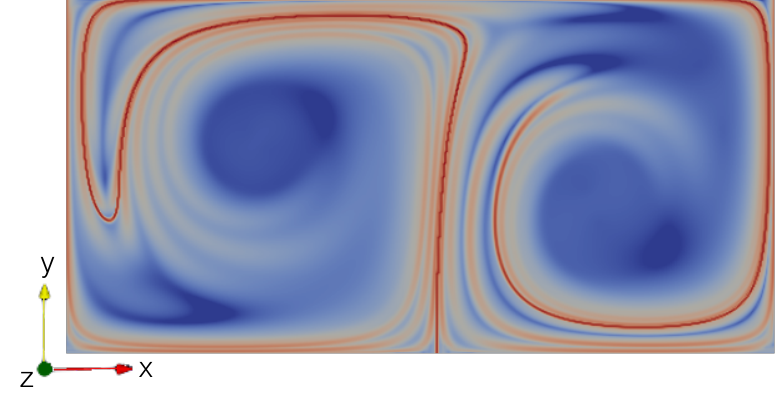}
    \caption{Forward-time FTLE field}
    \label{fig:DB1}
\end{subfigure}
\hspace{5mm}
\begin{subfigure}{0.47\textwidth}
    \includegraphics[width=\textwidth]{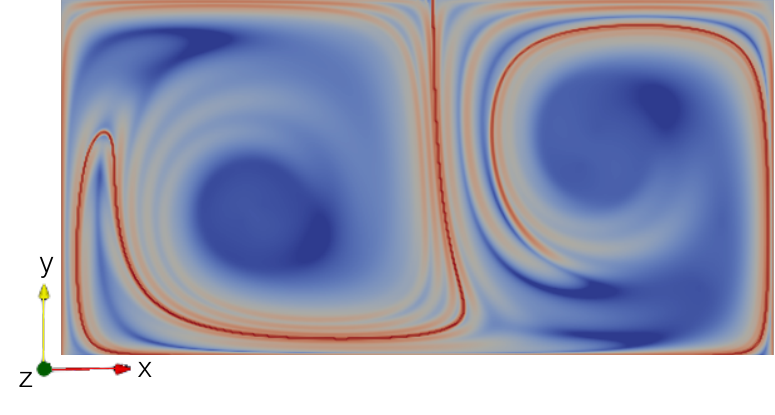}
    \caption{Backward-time FTLE field}
    \label{fig:DG2}
\end{subfigure}
\\
\begin{subfigure}{0.35\textwidth}
    \includegraphics[width=\textwidth]{figures/ABC_Legend.pdf}
\end{subfigure}
\caption{FTLE fields of a time-dependent double gyre at $t=15\unit{s}$.}
\label{fig:Double_gyre_results}
\end{figure}

Similar to the ABC flow example, a dedicated OpenFOAM solver was written for this case, which sets the velocity field on a given computational mesh according to (\ref{eqn:double_gyre_vel_comp}). For comparability, a mesh with the same specifications as in \cite{lipinski_ridge_2010},\cite{leung_eulerian_2011} and \cite{shadden_definition_2005} was used. It has the dimensions $\protect{[0,2]\times[0,1]\times[0,0.1]}m$ and a resolution of $512\times256\times1$ cells. As this mesh is also static and rectlinear no additional LCS mesh was used. For the mathematical model of the flow the parameter values are chosen to be $\epsilon=0.1$, $A=0.1 \unit{m}\,\unit{s}^{-1}$ and $\omega=2\pi/10 \unit{s}$. Since the oscillation frequency is known, the hydrodynamic time scale can be easily determined by $t_{ref}=2\pi/\omega=10 \unit{s}$. As described in section \ref{sec:Numerical_flow_map_computation}, the LCS integration time interval $T$ should be set larger than $t_{ref}$. Therefore, it is set to $T=1.5\cdot t_{ref}= 15 \unit{s}$. Figure \ref{fig:Double_gyre_results} shows the forward- and backward-time FTLE fields at $t= 15 \unit{s}$ of the previously described double gyre flow. Again, the results match very well with the results from \cite{lipinski_ridge_2010},\cite{leung_eulerian_2011} and \cite{shadden_definition_2005}. This confirms that the function object is able to calculate the correct FTLE fields from velocity fields generated by OpenFOAM.
    \subsection{Flow around cylinder}
        As it has already been shown in the previous examples that the function object can calculate the correct FTLE fields from velocity fields provided by OpenFOAM, this example will focus on how to deal with non-rectlinear simulation meshes. For this purpose, a standard flow problem is selected that is very well suited for an LCS evaluation: the flow around an infinitely long cylinder. 
The general case setup contains a fluid domain with size $\protect{[-20,30]\times[-20,20]\times[-0.5,0.5]}\unit{m}$ that surrounds a cylinder with diameter $D=2\unit{m}$ and its centre axis at $x=y=0\unit{m}$. The free-stream velocity and the fluids kinematic viscosity are set to $\mathbf{u}^{\sf T}=(1\;0\;0)\unit{m}\,\unit{s}^{-1}$ and $\nu = 0.01\unit{m}^{2}\unit{s}^{-1}$, respectively. This results in a Reynolds number of $\operatorname{Re}=200$ which indicates that vortex shedding behind the cylinder occurs in a barely laminar regime. If we also assume a Strouhal number of $\operatorname{St}=0.2$ at $\operatorname{Re}=200$, the hydrodynamic time scale of this flow is $t_{ref}=D/(u\cdot\operatorname{St})=10\unit{s}$.

Because of the cylinder in the middle of the domain, a computational mesh discretising this domain is no longer rectlinear. Therefore, we consider two different procedures in the LCS evaluation, the first of which is carried out in two different ways.

\begin{figure}[t]
\centering
\begin{subfigure}{0.35\textwidth}
    \includegraphics[width=\textwidth]{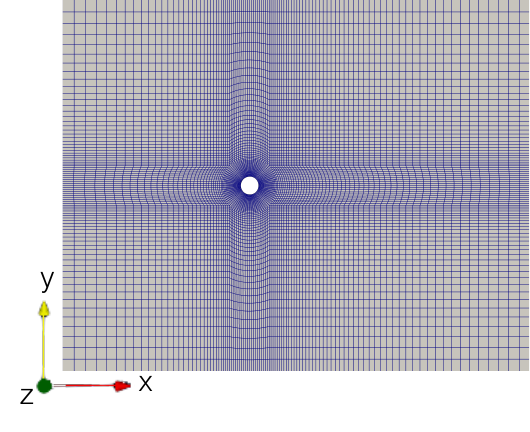}
    \caption{Non rectlinear simulation mesh \\ \hfill}
    \label{fig:CB1}
\end{subfigure}
\hspace{5mm}
\begin{subfigure}{0.35\textwidth}
    \includegraphics[width=\textwidth]{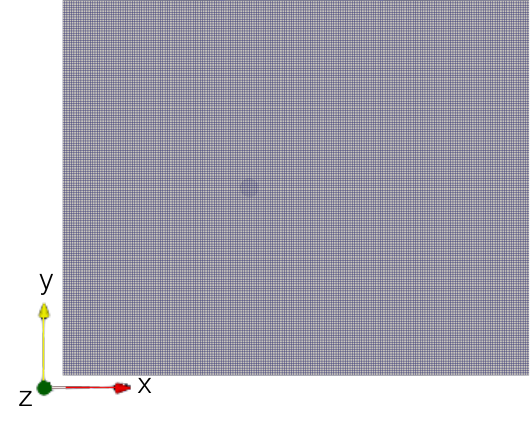}
    \caption{LCS mesh enclosing the whole domain}
    \label{fig:CB22}
\end{subfigure}
\\ \vspace{4mm}
\begin{subfigure}{0.35\textwidth}
    \includegraphics[width=\textwidth]{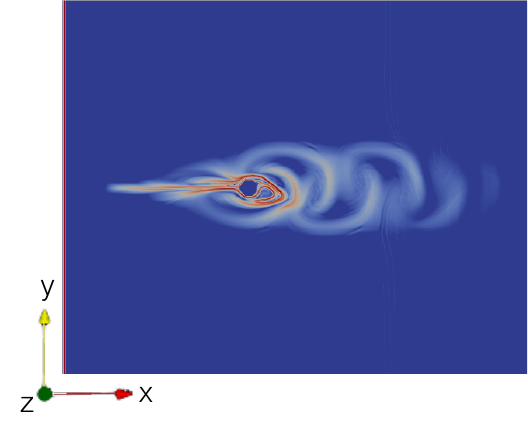}
    \caption{Forward-time FTLE field}
    \label{fig:CB3}
\end{subfigure}
\hspace{5mm}
\begin{subfigure}{0.35\textwidth}
    \includegraphics[width=\textwidth]{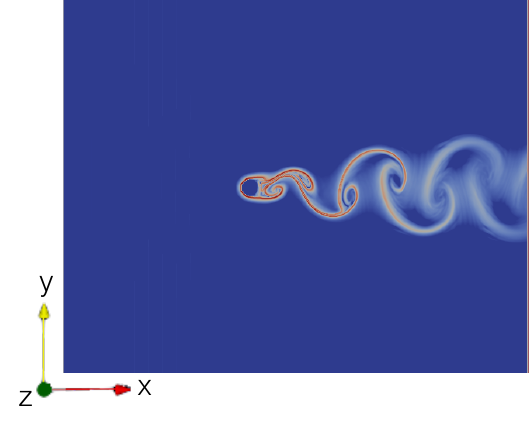}
    \caption{Backward-time FTLE field}
    \label{fig:CB4}
\end{subfigure}
\\
\begin{subfigure}{0.4\textwidth}
    \includegraphics[width=\textwidth]{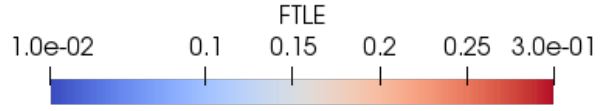}
\end{subfigure}
\caption{Simulation mesh, LCS mesh and FTLE fields of a flow around a cylinder with $\operatorname{Re}=200$ at $t=118\unit{s}$.}
\label{fig:Big_LCSMesh_results}
\end{figure}

Starting with the procedure where an additional rectlinear computational mesh is used for the LCS evaluation, the flow domain is discretised with a simulation mesh consisting of 9200 hexahedra (see upper left mesh in Fig.\ \ref{fig:Big_LCSMesh_results}).
The flow solver that is used to simulate the previously described flow from $t=0\unit{s}$ to $t=120\unit{s}$ is \texttt{pimpleFoam} with the initial conditions being calculated by \texttt{potentialFoam}. The first additional LCS mesh that is used within this procedure encloses the whole flow domain (see upper right mesh in Fig.\ \ref{fig:Big_LCSMesh_results}). In order to minimise the loss of information during the mapping of the velocity fields between the two grids, the resolution of the LCS mesh is chosen in a way that it corresponds approximately to the finest resolution in the simulation mesh. This leads to a LCS mesh with $200\times160\times1$ hexahedra. The boundary patch types are set to \texttt{patch} for the left and right patch (inlet,outlet), to \texttt{symmetryPlane} for the bottom and top patch and to \texttt{empty} for the front and back patch. The LCS integration time $T$ is again based on $t_{ref}$ and is set to $T=1.5\cdot t_{ref}=15\unit{s}$. For a good animation of the dynamics of the FTLE fields $h$ is chosen to be $h=T/10=1.5\unit{s}$. The results of the forward- and backward-time FTLE fields can be seen in Fig.\ \ref{fig:Big_LCSMesh_results}. They show how the vortices behind the cylinder form large coherent structures, where the FTLE ridges of the backward-time FTLE fields separate different fluid packages that do not mix in the vortex street.

\begin{figure}[t]
\centering
\begin{subfigure}{0.35\textwidth}
    \includegraphics[width=\textwidth]{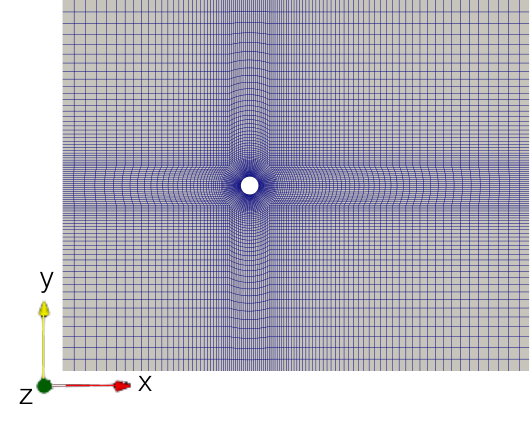}
    \caption{Non rectlinear simulation mesh \\ \hfill}
    \label{fig:CS1}
\end{subfigure}
\hspace{5mm}
\begin{subfigure}{0.35\textwidth}
    \includegraphics[width=\textwidth]{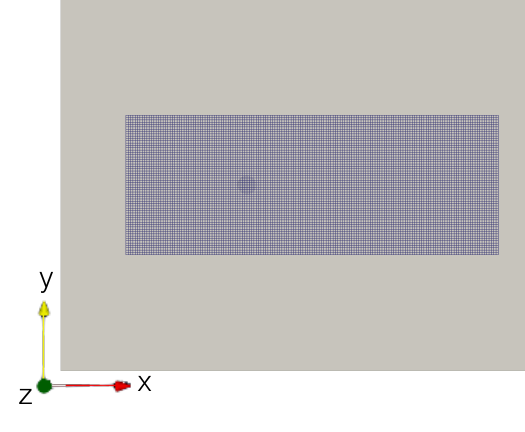}
    \caption{LCS mesh enclosing the region of interest}
    \label{fig:CS2}
\end{subfigure}
\\ \vspace{4mm}
\begin{subfigure}{0.35\textwidth}
    \includegraphics[width=\textwidth]{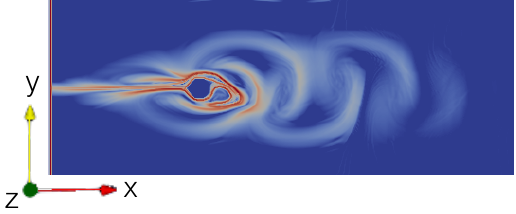}
    \caption{Forward-time FTLE field}
    \label{fig:CS3}
\end{subfigure}
\hspace{5mm}
\begin{subfigure}{0.35\textwidth}
    \includegraphics[width=\textwidth]{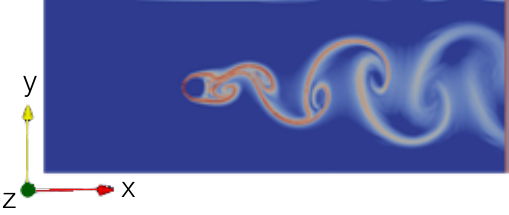}
    \caption{Backward-time FTLE field}
    \label{fig:CS4}
\end{subfigure}
\\
\begin{subfigure}{0.4\textwidth}
    \includegraphics[width=\textwidth]{figures/Cylinder_Legend.pdf}
\end{subfigure}
\caption{Simulation mesh, LCS mesh and FTLE fields of a flow around a cylinder with $\operatorname{Re}=200$ at $t=118\unit{s}$.}
\label{fig:Small_LCSMesh_results}
\end{figure}

Since the FTLE ridges only appear in a fraction of the overall domain and the LCS evaluation is computation-wise a quite costly operation, a second LCS mesh is prepared. This second LCS mesh is a lot smaller than the first one and encloses only the fraction of the flow domain where the FTLE ridges are expected to show up (see Fig.\ \ref{fig:Small_LCSMesh_results}). The boundary patches on the smaller LCS mesh and its spacial resolution are also set analogous to its bigger counterpart, leading to a LCS mesh of size $[-13,27]\times[-7.5,7.5]\times[-0.5,0.5]$ containing $160\times60\times1$ hexahedra. Repeating the computations with the use of the smaller LCS mesh gives the results which are displayed in Fig.\ \ref{fig:Small_LCSMesh_results} and are found to match with the results from the bigger LCS mesh. This shows that the LCS evaluation, when done with a separate LCS mesh, can be used in a very targeted way. The advantages this brings in terms of computational costs are discussed after considering the second procedure for the LCS evaluation of this flow problem.

\begin{figure}[t]
\centering
\begin{subfigure}{0.6\textwidth}
    \centering
    \includegraphics[width=0.6\textwidth]{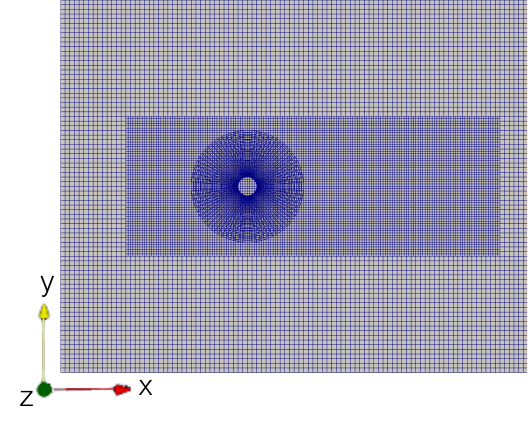}
    \caption{\texttt{oversetMesh} containing a rectlinear background mesh zone, a rectlinear refinement zone which is also used for the LCS evaluation, and a cylindrical mesh zone containing the cylinder.}
    \label{fig:CO1}
\end{subfigure}
\\ \vspace{4mm}
\begin{subfigure}{0.35\textwidth}
    \includegraphics[width=\textwidth]{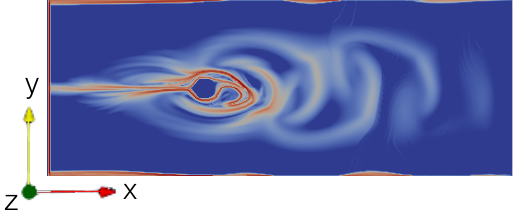}
    \caption{Forward-time FTLE field}
    \label{fig:CO3}
\end{subfigure}
\hspace{5mm}
\begin{subfigure}{0.35\textwidth}
    \includegraphics[width=\textwidth]{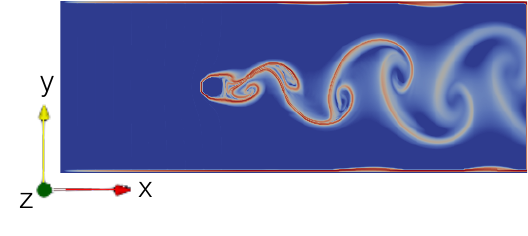}
    \caption{Backward-time FTLE field}
    \label{fig:CO4}
\end{subfigure}
\\
\begin{subfigure}{0.4\textwidth}
    \includegraphics[width=\textwidth]{figures/Cylinder_Legend.pdf}
\end{subfigure}
\caption{\texttt{oversetMesh} and FTLE fields of a flow around a cylinder with $\operatorname{Re}=200$ at $t=\unit[118]{s}$.}
\label{fig:OversetMesh_results}
\end{figure}

The second procedure, which can be used on problems where no single static rectlinear mesh can be constructed, utilises OpenFOAM's \texttt{oversetMesh} functionalities. With regard to the flow problem considered here, an \texttt{oversetMesh} is constructed with the same dimensions as the simulation mesh used previously. It consists of three mesh zones, namely a rectlinear background mesh zone that spans the whole fluid domain, another finer and smaller mesh zone that is used for a finer resolution of the flow and a cylindrical mesh that surrounds the cylinder (see Fig.\ \ref{fig:OversetMesh_results}). For comparability reasons the finer rectlinear mesh zone has the same dimensions and resolution as the smaller additional LCS mesh considered previously and is therefore specified as the cell zone for the LCS evaluation. Also all other LCS evaluation settings are adopted. The only difference to the previously considered simulations is the used flow solver. Here the flow solver is \texttt{pimpleDyMOversetFoam} due to the used \texttt{oversetMesh}. The resulting forward- and backward-time FTLE fields of this simulation can be found in Fig.\ \ref{fig:OversetMesh_results}. They match with the results from the previously considered procedure which shows that both approaches can be used equally well. The only thing that stands out are the high FTLE values along some boundaries in the studied solutions. These occur because of the way \textit{libcfd2lcs} handles its inlet and outlet boundary conditions. It fixes out-flowing Lagrangian particles/takeoff coordinates on "open" boundaries and cannot generate new in-flowing particles during the flow map computation. Therefore, high FTLE values occur in the forward-time FTLE fields at "open" boundaries where inflow occurs, since there the most "stretching" happens. Vice versa, high FTLE values occur in the backward-time FTLE fields at "open" boundaries where outflow occurs, since there the most "folding" happens. These high values at "open" boundaries are just artefacts and have to be neglected. The reason they appear more in the \texttt{oversetMesh} approach is that all \texttt{overset} type patches are passed to \textit{libcfd2lcs} as "open" boundaries whereas the user can specify all patches problem dependent in the additional LCS mesh approach.

Looking at the computation times of the flow calculations including the LCS evaluation, it becomes evident that LCS evaluation is a very costly operation (see Tab.\ \ref{tab:comp_times}). When using the "large" additional LCS mesh the simulation takes approximately 30 time longer than without the LCS evaluation. This can be improved by using the smaller additional LCS mesh. Here the simulation takes 9 times longer than without the LCS evaluation. Since the costs for the LCS evaluations are almost independent of the underlying simulation for a constant grid size, this factor becomes smaller and smaller for more complex simulations. This can also be seen from the fact that the factor is only 2.5 when the \texttt{oversetMesh} approach is used because the computations of the pressure and velocity fields take longer on an \texttt{oversetMesh}. At this point, however, it must be emphasised that the flow considered here is not a highly complex problem, which can also be seen from the simulation time of 1.5 min on a normal mesh and 8 min on an \texttt{oversetMesh}. 

\begin{table}
    \begin{center}
    \begin{tabular}{l  c  c  c }
    \toprule
    & "large" add. LCS mesh  & "small" add. LCS mesh & \texttt{oversetMesh} \\ \midrule
    Total time             & 44    & 13.5  & 20 \\  
    Simulation time        & 1.5   & 1.5   & 8 \\ 
    Total FTLE comp. time  & 42.5  & 12    & 12 \\
    Fwd.-FTLE comp. time   & 8.5   & 2     & 2 \\
    Bkwd.-FTLE comp. time  & 21    & 6     & 6 \\
    I/O, flow map const., ... & 13 & 4 & 4\\
    \bottomrule
    \end{tabular}
    \caption{Computing times in minutes for the flow simulation including LCS evaluation on 2 cores of a Intel Core i5 processor}
    \label{tab:comp_times}
\end{center}
\end{table}

\section{Summary \& Conclusion}
    We provide an OpenFOAM function object based on \textit{libcfd2lcs} to compute Finite-Time Lyapunov Exponent (FTLE) fields that indicate candidates of Lagrangian Coherent Structures (LCS) and allow to visualise finite-time stretching and folding fields. LCS reveal the robust skeleton of material surfaces and are key to quantitatively assess material transport in time-dependent flows. This enables the OpenFOAM community to assess the geometry of the material transport in any flow quantitatively on-the-fly using principally any OpenFOAM flow solver.

Focusing on the practical aspects, we only give a brief overview of the mathematical foundation as well as how the computation is done in practice. We describe the structure and functionality of the newly developed function object. Further focus is laid on how the function object acts as an interface between OpenFOAM and \textit{libcfd2lcs}, how parallelisation is ensured and what has to be considered for the output of the generated data.

From validation of the presented function object using simple benchmark problems, a notable computational overhead has been recognised. However, if LCS evaluations are used for much more complex problems as the ones used here, the computational overhead significantly drops and the LCS evaluation no longer accounts for the largest proportion of the computation time. Nevertheless, the user should be aware that the calculation of FTLE fields is expensive and should therefore think carefully about the size and position of the LCS mesh. In addition, consideration should also be given to whether both forward and backward-time FTLE calculations are required or if one of them is sufficient.


%





%
%

\bibliographystyle{IEEEtran}

\bibliography{LCSFunctionObject}

\begin{thebibliography}{10}
\providecommand{\url}[1]{#1}
\csname url@samestyle\endcsname
\providecommand{\newblock}{\relax}
\providecommand{\bibinfo}[2]{#2}
\providecommand{\BIBentrySTDinterwordspacing}{\spaceskip=0pt\relax}
\providecommand{\BIBentryALTinterwordstretchfactor}{4}
\providecommand{\BIBentryALTinterwordspacing}{\spaceskip=\fontdimen2\font plus
\BIBentryALTinterwordstretchfactor\fontdimen3\font minus
  \fontdimen4\font\relax}
\providecommand{\BIBforeignlanguage}[2]{{%
\expandafter\ifx\csname l@#1\endcsname\relax
\typeout{** WARNING: IEEEtran.bst: No hyphenation pattern has been}%
\typeout{** loaded for the language `#1'. Using the pattern for}%
\typeout{** the default language instead.}%
\else
\language=\csname l@#1\endcsname
\fi
#2}}
\providecommand{\BIBdecl}{\relax}
\BIBdecl

\bibitem{kameke_how_2019}
\BIBentryALTinterwordspacing
A.~Kameke, S.~Kastens, S.~Rüttinger, S.~Herres-Pawlis, and M.~Schlüter,
  ``\BIBforeignlanguage{en}{How coherent structures dominate the residence time
  in a bubble wake: {An} experimental example},''
  \emph{\BIBforeignlanguage{en}{Chemical Engineering Science}}, vol. 207, pp.
  317--326, Nov. 2019. [Online]. Available:
  \url{https://linkinghub.elsevier.com/retrieve/pii/S0009250919305366}
\BIBentrySTDinterwordspacing

\bibitem{huhn_impact_2012}
\BIBentryALTinterwordspacing
F.~Huhn, A.~von Kameke, V.~Pérez-Muñuzuri, M.~J. Olascoaga, and F.~J.
  Beron-Vera, ``\BIBforeignlanguage{en}{The impact of advective transport by
  the {South} {Indian} {Ocean} {Countercurrent} on the {Madagascar} plankton
  bloom: {Jet} in {Madagascar} {Plankton} {Bloom}},''
  \emph{\BIBforeignlanguage{en}{Geophysical Research Letters}}, vol.~39, no.~6,
  pp. n/a--n/a, Mar. 2012. [Online]. Available:
  \url{http://doi.wiley.com/10.1029/2012GL051246}
\BIBentrySTDinterwordspacing

\bibitem{dovidio_mixing_2004}
\BIBentryALTinterwordspacing
F.~d'Ovidio, V.~Fernández, E.~Hernández-García, and C.~López,
  ``\BIBforeignlanguage{en}{Mixing structures in the {Mediterranean} {Sea} from
  finite-size {Lyapunov} exponents: {Mixing} {Structures} in the
  {Mediterranean} {Sea}},'' \emph{\BIBforeignlanguage{en}{Geophysical Research
  Letters}}, vol.~31, no.~17, pp. n/a--n/a, Sep. 2004. [Online]. Available:
  \url{http://doi.wiley.com/10.1029/2004GL020328}
\BIBentrySTDinterwordspacing

\bibitem{weiss_transport_2008}
\BIBentryALTinterwordspacing
J.~B. Weiss and A.~Provenzale, Eds., \emph{\BIBforeignlanguage{en}{Transport
  and {Mixing} in {Geophysical} {Flows}: {Creators} of {Modern}
  {Physics}}}.\hskip 1em plus 0.5em minus 0.4em\relax Berlin, Heidelberg:
  Springer Berlin Heidelberg, 2008. [Online]. Available:
  \url{http://link.springer.com/10.1007/978-3-540-75215-8}
\BIBentrySTDinterwordspacing

\bibitem{neufeld_chemical_2009}
\BIBentryALTinterwordspacing
Z.~Neufeld and E.~Hernández-García, \emph{\BIBforeignlanguage{en}{Chemical
  and {Biological} {Processes} in {Fluid} {Flows}: {A} {Dynamical} {Systems}
  {Approach}}}.\hskip 1em plus 0.5em minus 0.4em\relax Imperial College Press,
  Sep. 2009. [Online]. Available:
  \url{https://www.worldscientific.com/worldscibooks/10.1142/p471}
\BIBentrySTDinterwordspacing

\bibitem{balasuriya_generalized_2018}
\BIBentryALTinterwordspacing
S.~Balasuriya, N.~T. Ouellette, and I.~I. Rypina,
  ``\BIBforeignlanguage{en}{Generalized {Lagrangian} coherent structures},''
  \emph{\BIBforeignlanguage{en}{Physica D: Nonlinear Phenomena}}, vol. 372, pp.
  31--51, Jun. 2018. [Online]. Available:
  \url{https://linkinghub.elsevier.com/retrieve/pii/S0167278917302750}
\BIBentrySTDinterwordspacing

\bibitem{hadjighasem_critical_2017}
\BIBentryALTinterwordspacing
A.~Hadjighasem, M.~Farazmand, D.~Blazevski, G.~Froyland, and G.~Haller,
  ``\BIBforeignlanguage{en}{A critical comparison of {Lagrangian} methods for
  coherent structure detection},'' \emph{\BIBforeignlanguage{en}{Chaos: An
  Interdisciplinary Journal of Nonlinear Science}}, vol.~27, no.~5, p. 053104,
  May 2017. [Online]. Available:
  \url{http://aip.scitation.org/doi/10.1063/1.4982720}
\BIBentrySTDinterwordspacing

\bibitem{justin_finn_libcfd2lcs_2017}
\BIBentryALTinterwordspacing
{Justin Finn}, {Romain Watteaux}, and {Andrew Lawrie},
  ``\BIBforeignlanguage{English}{libcfd2lcs: {A} general purpose library for
  computation of {Lagrangian} coherent structures during {CFD} simulation},''
  Feb. 2017. [Online]. Available:
  \url{https://www.archer.ac.uk/training/virtual/2016-06-22-libcfd2lcs/ArcherVT.pdf}
\BIBentrySTDinterwordspacing

\bibitem{finn_integrated_2013}
\BIBentryALTinterwordspacing
J.~Finn and S.~V. Apte, ``\BIBforeignlanguage{en}{Integrated computation of
  finite-time {Lyapunov} exponent fields during direct numerical simulation of
  unsteady flows},'' \emph{\BIBforeignlanguage{en}{Chaos: An Interdisciplinary
  Journal of Nonlinear Science}}, vol.~23, no.~1, p. 013145, Mar. 2013.
  [Online]. Available: \url{http://aip.scitation.org/doi/10.1063/1.4795749}
\BIBentrySTDinterwordspacing

\bibitem{farazmand_computing_2012}
\BIBentryALTinterwordspacing
M.~Farazmand and G.~Haller, ``\BIBforeignlanguage{en}{Computing {Lagrangian}
  coherent structures from their variational theory},''
  \emph{\BIBforeignlanguage{en}{Chaos: An Interdisciplinary Journal of
  Nonlinear Science}}, vol.~22, no.~1, p. 013128, Mar. 2012. [Online].
  Available: \url{http://aip.scitation.org/doi/10.1063/1.3690153}
\BIBentrySTDinterwordspacing

\bibitem{shadden_definition_2005}
\BIBentryALTinterwordspacing
S.~C. Shadden, F.~Lekien, and J.~E. Marsden,
  ``\BIBforeignlanguage{en}{Definition and properties of {Lagrangian} coherent
  structures from finite-time {Lyapunov} exponents in two-dimensional aperiodic
  flows},'' \emph{\BIBforeignlanguage{en}{Physica D: Nonlinear Phenomena}},
  vol. 212, no. 3-4, pp. 271--304, Dec. 2005. [Online]. Available:
  \url{https://linkinghub.elsevier.com/retrieve/pii/S0167278905004446}
\BIBentrySTDinterwordspacing

\bibitem{haller_lagrangian_2015}
\BIBentryALTinterwordspacing
G.~Haller, ``\BIBforeignlanguage{en}{Lagrangian {Coherent} {Structures}},''
  \emph{\BIBforeignlanguage{en}{Annual Review of Fluid Mechanics}}, vol.~47,
  no.~1, pp. 137--162, Jan. 2015. [Online]. Available:
  \url{https://www.annualreviews.org/doi/10.1146/annurev-fluid-010313-141322}
\BIBentrySTDinterwordspacing

\bibitem{brunton_fast_2010}
\BIBentryALTinterwordspacing
S.~L. Brunton and C.~W. Rowley, ``\BIBforeignlanguage{en}{Fast computation of
  finite-time {Lyapunov} exponent fields for unsteady flows},''
  \emph{\BIBforeignlanguage{en}{Chaos: An Interdisciplinary Journal of
  Nonlinear Science}}, vol.~20, no.~1, p. 017503, Mar. 2010. [Online].
  Available: \url{http://aip.scitation.org/doi/10.1063/1.3270044}
\BIBentrySTDinterwordspacing

\bibitem{leung_eulerian_2011}
\BIBentryALTinterwordspacing
S.~Leung, ``\BIBforeignlanguage{en}{An {Eulerian} approach for computing the
  finite time {Lyapunov} exponent},'' \emph{\BIBforeignlanguage{en}{Journal of
  Computational Physics}}, vol. 230, no.~9, pp. 3500--3524, May 2011. [Online].
  Available:
  \url{https://linkinghub.elsevier.com/retrieve/pii/S0021999111000799}
\BIBentrySTDinterwordspacing

\bibitem{haller_lagrangian_2002}
\BIBentryALTinterwordspacing
G.~Haller, ``\BIBforeignlanguage{en}{Lagrangian coherent structures from
  approximate velocity data},'' \emph{\BIBforeignlanguage{en}{Physics of
  Fluids}}, vol.~14, no.~6, pp. 1851--1861, Jun. 2002. [Online]. Available:
  \url{http://aip.scitation.org/doi/10.1063/1.1477449}
\BIBentrySTDinterwordspacing

\bibitem{sulman_leaving_2013}
\BIBentryALTinterwordspacing
M.~H. Sulman, H.~S. Huntley, B.~Lipphardt, and A.~Kirwan,
  ``\BIBforeignlanguage{en}{Leaving flatland: {Diagnostics} for {Lagrangian}
  coherent structures in three-dimensional flows},''
  \emph{\BIBforeignlanguage{en}{Physica D: Nonlinear Phenomena}}, vol. 258, pp.
  77--92, Sep. 2013. [Online]. Available:
  \url{https://linkinghub.elsevier.com/retrieve/pii/S0167278913001450}
\BIBentrySTDinterwordspacing

\bibitem{solomon_chaotic_1989}
T.~Solomon and J.~Gollub, ``Chaotic {Transport} in {Time}-{Dependent}
  {Rayleigh}-{Bénard} {Convection},'' \emph{Physical review. A}, vol.~38, pp.
  6280--6286, Jan. 1989.

\bibitem{lipinski_ridge_2010}
\BIBentryALTinterwordspacing
D.~Lipinski and K.~Mohseni, ``\BIBforeignlanguage{en}{A ridge tracking
  algorithm and error estimate for efficient computation of {Lagrangian}
  coherent structures},'' \emph{\BIBforeignlanguage{en}{Chaos: An
  Interdisciplinary Journal of Nonlinear Science}}, vol.~20, no.~1, p. 017504,
  Mar. 2010. [Online]. Available:
  \url{http://aip.scitation.org/doi/10.1063/1.3270049}
\BIBentrySTDinterwordspacing

\end{thebibliography}

\end{document}